\documentstyle[amssymb,twocolumn,aps]{revtex}
\begin{document}
\draft
\title{Saddle point states and energy barriers for vortex entrance and exit in
superconducting disks and rings}
\author{B. J. Baelus, F. M. Peeters\cite{peeters} and V. A. Schweigert\cite
{schweigert}}
\address{Departement Natuurkunde, Universiteit Antwerpen (UIA), Universiteitsplein 1,%
\\
B-2610 Antwerpen, Belgium}
\date{\today}
\maketitle

\begin{abstract}
The transitions between the different vortex states of thin mesoscopic
superconducting disks and rings are studied using the non-linear
Ginzburg-Landau functional. They are saddle points of the free energy
representing the energy barrier which has to be overcome for transition
between the different vortex states. In small superconducting disks and
rings the saddle point state between two giant vortex states, and in larger
systems the saddle point state between a multivortex state and a giant
vortex state and between two multivortex states is obtained. The shape and
the height of the nucleation barrier is investigated for different disk and
ring configurations.
\end{abstract}

\pacs{74.60.Ec, 74.20.De, 73.23.-b}

\section{Introduction}

The study of superconducting samples with sizes comparable to the
penetration depth $\left( \lambda \right) $ and the coherence length $\left(
\xi \right) $ became possible due to recent progress in nanofabrication
technologies. This evolution resulted in an increase of interest in the
investigation of flux penetration and flux expulsion in such mesoscopic
samples in order to explain the hysteresis behavior and the different phase
transitions in thin superconducting samples \cite
{Buisson,Geim,PRL79,Benoist,PRB57,PRL81,Pal1,Fomin,Akkermans,SM25,Bruyndoncx2,Berger,Bruyndoncx,PRB61,PC322p}%
.

It is well known that for type-II $\left( \kappa =\lambda /\xi >1/\sqrt{2}%
\right) $ superconductors the triangular Abrikosov vortex lattice is
energetically favored in the magnetic field range $H_{c1}<H<H_{c2}$ where $%
\kappa $ is the Ginzburg-Landau (GL) parameter, and $H_{c1}$ and $H_{c2}$
are the first and second critical fields of a Type-II superconductor. Since
the effective London penetration depth $\Lambda =\lambda ^{2}/d$ increases
considerably in thin films one expects the appearance of the Abrikosov
multivortex state even in thin Type-I $\left( \kappa <1/\sqrt{2}\right) $
superconductors when the thickness $d\ll \lambda $. But, in small confined
systems there is a competition between the boundary of the sample, which
tries to impose the symmetry of the sample boundary on the vortex
configuration, and this triangular Abrikosov state. As a consequence, the
effective GL parameter $\kappa ^{\ast }=\Lambda /\xi $ is no longer the only
controlling parameter which determines the shape of the vortex configuration
in thin mesocopic superconducting samples \cite{PRL81}. Previous theoretical
and experimental studies of superconducting disks and rings \cite
{Buisson,Geim,PRL79,Benoist,PRB57,PRL81,Pal1,Fomin,Akkermans,SM25,Bruyndoncx2,Berger,Bruyndoncx,PRB61,PC322p,Fink}
found that, as function of the applied field, there are transitions between
circular symmetric vortex states (called giant vortex states) with different
vorticity $L$. Experimentally it was found that the magnetic field at which
the transition $L\rightarrow L+1$ occurs does not necessarily coincides with
the magnetic field $H_{tr}$ where the vorticity of the ground state changes
from $L$ to $L+1$, i.e. it is possible to drive the system in a metastable
state. This is typical for first order phase transitions. For increasing
applied field, the state with vorticity $L$ remains stable up to the
penetration field $H_{p}>H_{tr}$ and transits then to the superconducting
state with vorticity $L+1$. For decreasing applied field, the state with
vorticity $L+1$ remains stable down to the expulsion field $H_{e}<H_{tr}$
before going to the state with vorticity $L$. This hysteresis effect is a
consequence of the presence of an energy barrier between the states with
vorticity $L$ and $L+1$. The latter corresponds to different minima of the
free energy in configurational space and the lowest barrier between those
two minima is a saddle point. This barrier arises from the fact that the
superconducting current around a vortex is in the opposite direction to the
screening currents at the surface of the sample \cite{BL}. This
Bean-Livingston model has been refined to different sample geometries \cite
{Galaiko,Fetter,Kogan,Indenbom,Langer,PRB59}. The time of flux penetration
and expulsion is determined by the height of the energy barrier.

The experimental consequences of the existence of these metastable states
are: (i) hysteretic behavior \cite{Geim}, (ii) paramagnetic Meissner effect 
\cite{Geim2,Akkermans2,Pal3,PC332,Moshchalkov,Zharkov}, (iii) fractional
flux penetration \cite{Geim3}, and (iv) negative flux entrance \cite{Geim3},
i.e. a decrease of the flux penetration through the superconducting disk
with increasing vorticity and increasing magnetic field.

Schweigert and Peeters \cite{PRL83} studied flux penetration and expulsion
in thin superconducting disks and presented an approach to find the saddle
point states. They calculated the height of the free energy barriers which
separate the stable states with different vorticity $L.$ We will extend
their approach and present a systematic study of flux penetration and
expulsion in thin superconducting disks and disks with a hole in the center,
i.e. mesoscopic ring structures.

Bezryadin {\it et al} \cite{Bezryadin}\ used the nonlinear GL equation to
study the phase diagram of a thin-wire loop and a thin film with a circular
hole in the limit $\kappa ^{\ast }\gg 1$. They performed a stability
analysis of the giant vortex state with vorticity $L$ by allowing only the
admixture of the $L+1$ vortex state. A more rigorous stability analysis was
performed by Horane {\it et al} \cite{Horane}\ who studied the saddle points
between two vortex states of a one dimensional wire of zero width. They
allowed for more possible non-uniform perturbations which may make the
vortex state unstable. They found that the transition between two angular
momentum states occurs through a saddle point which has a zero in the order
parameter at some point along the ring. Such a zero creates a phase slip
center, allowing the phase winding required for the transition. Our systems
have a non zero radial width and consequently such a scenario is not
possible because the order parameter is not allowed to be zero along a
radial line. In fact it was found in Ref.~\cite{PRL83} that for a disk
geometry, the saddle point for flux penetration corresponds to a state with
suppressed superconductivity at the disk edge which acts as a nucleus for
the following vortex creation. In the present paper we will find that for
rings with a finite width this picture has to be modified because of the
presence of two boundaries, i.e. two edges.

Recently, Palacios \cite{Pal3} calculated saddle point states and the energy
barriers responsible for the metastabilities of superconducting mesoscopic
disks using the lowest Landau level approximation. The central idea of his
method was to find generic stationary solutions of the Ginzburg-Landau
functional and to project the order parameter onto smaller subspaces spanned
by a finite number $l$ of eigenfunctions, $\left\{
L_{1},L_{2},...,L_{l}\right\} $, where $0\leq L_{1}\leq L_{2}\leq ...\leq
L_{l}$. Palacios restricted himself to $l\leq 3$ and therefore his approach
is a special case of the one of Ref.~\cite{PRL83} where no such restriction
on $l$ was imposed and where also different radial states were included.
Yampolskii and Peeters \cite{PRB62} investigated the influence of the
boundary condition (surface enhancement) on the superconducting states and
the energy barriers between those vortex states. They also restricted their
calculations to $l\leq 3$.

Recently, Akkermans {\it et al} \cite{Akkermans2} studied the behavior of
metastable vortex states in infinite superconducting cylinders for $\kappa
\gg 1$, i.e. the London regime. They considered the situation where the
vortices are symmetrically distributed along a closed ring and they found
structural phase transitions of vortex patterns between the metastable
states. The key concept was the introduction of a special curve $\Gamma $,
which embodies the main geometric features of a vortex configuration. This
curve appears mathematically as a limit cycle of the system of currents
generated by the vortex pattern and separates the paramagnetic and
diamagnetic domains.

The paper is organized as follows: In Sec.~II we present the theoretical
model and the calculation method to obtain the saddle points. In Sec.~III we
study thin superconducting disks and extend and supplement our previous
results \cite{PRL83}. We make a distinction between small and large
superconducting disks. In small disks only the giant vortex state appears,
while in larger disks multivortices can nucleate and transitions between
different multivortices are possible \cite{Zalalutdinov,Geim4}. In Sec.~IV
we consider superconducting rings, where we make a distinction between small
and large rings. Our results are summarized in Sec.~V.

\section{Theoretical formalism}

In the present paper we consider very thin superconducting disks with radius 
$R$ and thickness $d$, and superconducting rings with inner radius $R_{i}$
and outer radius $R_{o}$. These mesoscopic superconducting systems are
immersed in an insulating medium in the presence of a perpendicular uniform
magnetic field $H_{0}$. To solve this problem, we follow the numerical
approach of Schweigert and Peeters \cite{PRL83}. For very thin disks and
rings, i.e. $Wd\ll \lambda ^{2}$, with $W=R$ the radius of the disk or $%
W=R_{o}-R_{i}$ the width of the ring, the demagnetization effects can be
neglected and the Ginzburg-Landau functional can be written as 
\begin{equation}
G=G_{n}+\int d\overrightarrow{r}\left( \alpha \left| \Psi \right| ^{2}+\frac{%
\beta }{2}\left| \Psi \right| ^{4}+\Psi ^{\ast }\widehat{L}\Psi \right) 
\text{,}
\end{equation}
where $G,$ $G_{n}$ are the free energies of the superconducting and the
normal states, $\Psi $ is the complex order parameter, $\alpha $ and $\beta $
are the GL coefficients which depend on the sample temperature. $\widehat{L}$
is the kinetic energy operator for Cooper-pairs of charge $e^{\ast }=2e$ and
mass $m^{\ast }=2m$, i.e. 
\begin{equation}
\widehat{L}=(-i\text{{\it 
%TCIMACRO{\UNICODE{0x127}}%
%BeginExpansion
h\hskip-.2em\llap{\protect\rule[1.1ex]{.325em}{.1ex}}\hskip.2em%
%EndExpansion
}}\overrightarrow{\nabla }-e^{\ast }\overrightarrow{A}/c)/2m^{\ast }\text{,}
\label{vgl2}
\end{equation}
where $\overrightarrow{A}=\overrightarrow{e}_{\phi }H_{0}\rho /2$ is the
vector potential of the uniform magnetic field $H_{0}$ written in
cylindrical coordinates $\rho $ and $\phi $.

By expanding the order parameter $\Psi =\sum_{i}^{N}C_{i}\varphi _{i}$ in
the orthonormal eigenfunctions of the kinetic energy operator $\widehat{L}%
\varphi _{i}=\epsilon _{i}\varphi _{i}$ \cite{PRB57,PRL81,Pal1}, the
difference between the superconducting and the normal state Gibbs free
energy can be written in terms of complex variables as 
\begin{equation}
F=G-G_{n}=\left( \alpha +\epsilon _{i}\right) C_{i}C_{i}^{\ast }+\frac{\beta 
}{2}A_{kl}^{ij}C_{i}^{\ast }C_{j}^{\ast }C_{k}C_{l}\text{,}  \label{vgl3}
\end{equation}
where the matrix elements $A_{kl}^{ij}=\int d\overrightarrow{r}\varphi
_{i}^{\ast }\varphi _{j}^{\ast }\varphi _{k}\varphi _{l}$ are calculated
numerically. The boundary condition for these $\varphi _{i}$, corresponding
to zero current density in the insulator media, is 
\begin{equation}
\left. \left( -i\text{{\it 
%TCIMACRO{\UNICODE{0x127}}%
%BeginExpansion
h\hskip-.2em\llap{\protect\rule[1.1ex]{.325em}{.1ex}}\hskip.2em%
%EndExpansion
}}\overrightarrow{\nabla }-\frac{e^{\ast }\overrightarrow{A}}{c}\right)
\right| _{n}\varphi _{i}=0\text{ .}
\end{equation}
These eigenenergies $\epsilon _{i}$ and the eigenfunctions $\varphi _{i}$
depend on the sample geometry. For thin axial symmetric samples the
eigenfunctions have the form $\varphi _{j=(n,l)}(\rho ,\phi )=exp(il\phi
)f_{n}(\rho )$, where $l$ is the angular momentum and the index $n$ counts
different states with the same $l$. Thus, the order parameter $\Psi $ can be
written as 
\begin{equation}
\psi =\sum_{n}\sum_{l}C_{n,l}\varphi _{n,l}\text{ .}  \label{vgl4}
\end{equation}
We do not restrict ourselves to the lowest landau level approximation (i.e. $%
n=1$) and expand the order parameter over all eigenfunctions with energy $%
\epsilon _{i}<\epsilon _{\ast }$, where the cutting parameter $\epsilon
_{\ast }$ is chosen such that increasing it does not influence the results.
The typical number of complex components used are in the range $N=30-50$.
Thus the superconducting state is mapped into a 2D cluster of $N$ particles
with coordinates $(x_{i},y_{i})\leftrightarrow (%
%TCIMACRO{\func{Re}}%
%BeginExpansion
\mathop{\rm Re}%
%EndExpansion
(C_{i}),%
%TCIMACRO{\func{Im}}%
%BeginExpansion
\mathop{\rm Im}%
%EndExpansion
(C_{i})),$ whose energy is determined by the Hamiltonian (\ref{vgl3}). The
energy landscape in this $2N+1$ dimensional space is studied where the local
minima and the saddle points between them will be determined together with
the corresponding vortex states.

To find the superconducting states and the saddle point states we use the
technique described in Ref.~\cite{PRL83}. A particular state is given by its
set of coefficients $\left\{ C_{i}\right\} $. We calculate the free energy
in the vicinity of this point $\delta G=G(C^{n})-G(C)$ where $\left\{
C^{n}\right\} $ is the set of coefficients of a state very close to the
initial one. This free energy is expanded to second order in the deviations $%
\delta =C^{n}-C,$%
\begin{equation}
\delta G=F_{m}\delta _{m}^{\ast }+B_{mn}\delta _{n}\delta _{m}^{\ast
}+D_{mn}\delta _{n}^{\ast }\delta _{m}^{\ast }+c.c.\text{ ,}
\end{equation}
where 
\begin{eqnarray}
F_{m} &=&\left( \alpha +\epsilon _{i}\right) C_{m}+\beta
A_{kl}^{mj}C_{j}C_{k}^{\ast }C_{l}\text{ ,}  \eqnum{7a} \\
B_{mn} &=&\left( \alpha +\epsilon _{m}\right) I_{mn}+2\beta
A_{kl}^{mn}C_{k}C_{l}^{\ast }\text{ ,}  \eqnum{7b} \\
D_{mn} &=&\beta A_{kl}^{mn}C_{k}C_{l}\text{ ,}  \eqnum{7c}
\end{eqnarray}
and $I_{mn}$ is the unit matrix. Using normal coordinates $\delta
_{m}=x_{k}Q_{m}^{k}$ we can rewrite the quadratic form as $\delta G=2\left(
\gamma _{k}x_{k}+\eta _{k}x_{k}^{2}\right) .$ To find the eigenvalues $\eta
_{k}$ and the eigenvectors $\gamma _{k}$ we solve numerically the following
equation: 
\begin{equation}
\left| 
\begin{array}{cc}
B+%
%TCIMACRO{\func{Re}}%
%BeginExpansion
\mathop{\rm Re}%
%EndExpansion
(D) & 
%TCIMACRO{\func{Im}}%
%BeginExpansion
\mathop{\rm Im}%
%EndExpansion
(D) \\ 
%TCIMACRO{\func{Im}}%
%BeginExpansion
\mathop{\rm Im}%
%EndExpansion
(D) & B-%
%TCIMACRO{\func{Re}}%
%BeginExpansion
\mathop{\rm Re}%
%EndExpansion
(D)
\end{array}
\right| .\left| 
\begin{array}{c}
%TCIMACRO{\func{Re}}%
%BeginExpansion
\mathop{\rm Re}%
%EndExpansion
(Q^{k}) \\ 
%TCIMACRO{\func{Im}}%
%BeginExpansion
\mathop{\rm Im}%
%EndExpansion
(Q^{k})
\end{array}
\right| =\eta _{k}\left| 
\begin{array}{c}
%TCIMACRO{\func{Re}}%
%BeginExpansion
\mathop{\rm Re}%
%EndExpansion
(Q^{k}) \\ 
%TCIMACRO{\func{Im}}%
%BeginExpansion
\mathop{\rm Im}%
%EndExpansion
(Q^{k})
\end{array}
\right| \text{ .}  \eqnum{8}
\end{equation}
Starting from a randomly choosen initial set of coefficients, we calculate a
nearby minimum of the free energy by moving in the direction of the negative
free energy gradient $-\gamma _{k}$. The set of coefficients of this minimum
determines then the ground state or a metastable state. Starting from the
initial set of coefficients we can also calculate a nearby saddle point
state by moving to a minimum of the free energy in all directions, except
the one which has the lowest eigenvalue. In this direction we move to a
local maximum. Repeating this procedure for many randomly choosen initial
sets of coefficients $\left\{ C_{i}\right\} $ for fixed magnetic field, we
find the different possible superconducting states and saddle point states.
To calculate the magnetic field dependence we start from a superconducting
state at a certain field and we change the applied field by small
increments. By moving into the direction of the nearest minimum or saddle
point, the corresponding state will be found for the new magnetic field,
provided that the field step is small enough.

\section{Superconducting disks}

In the present section we discuss superconducting disks. Although the system
is circular symmetric, in general we are not allowed to assume that $\Psi
(\rho ,\phi )=F\left( \rho \right) e^{iL\phi }$ because of the non-linear
term in the GL functional. Nevertheless, in small disks the confinement
effects are dominant and this imposes a circular symmetry on the
superconducting condensate, which means that only the `giant' vortex state,
i.e. a circular symmetric vortex state, appears. For larger disks and not
too large magnetic fields, the confinement effects are no longer dominant
and multivortices can nucleate in a certain magnetic field range. For this
reason we make a distinction between small and large disks.

\subsection{Small disks: giant vortex state}

We consider superconducting disks with radius $R=2.0\xi $. First we
investigate the influence of the number of terms in the expansion of Eq.~(%
\ref{vgl4}) on the energy of the minima and the saddle points. For the
approach that $n=1$ (i.e. lowest Landau level) and if only one $l$ is taken
into account for each state, i.e. $\Psi =C_{l}\varphi _{l}$, we find three
different states for three different values of $l$, $l=0$, $1$ and $2$, in
different magnetic field regions. In Fig.~\ref{enR2}(a) the free energy $F$\
of these $L$-states, measured in the condensation energy $F_{0}=\alpha
^{2}\pi R^{2}d/2\beta $, is shown by the dotted curves as a function of the
applied magnetic field $H_{0}$. Next, we take into account two values of $l$
and $n=1$, i.e. $\Psi =C_{l_{1}}\varphi _{l_{1}}+C_{l_{2}}\varphi _{l_{2}}$
as was done in e.g. Ref.~\cite{Pal3}. With this approach we find $L$-states
with $L=0,1,2$ and because of the concomittant existence of two minima also
saddle point states with $\left( l_{1},l_{2}\right) =(0,1)$ and $(1,2)$
appear. These are the saddle point states for the transition between $L$%
-states with $L=l_{1}$ and $L=l_{2}$. In Fig.~\ref{enR2}(a) the $L$-states
for this approach are given by the solid curves and the saddle point states
by the dashed curves. The inset shows the transition between the $L=1$ state
and the $L=2$ state in more detail. Notice that including one extra term in
Eq.~(\ref{vgl4}) reduces appreciably the stability region of the different
giant vortex states, i.e. its metastable region is strongly reduced. The $L$%
-states are only stable up to the point where its energy equals the saddle
point states.

Fig.~\ref{enR2}(b) shows the free energy as a function of the applied
magnetic field if we do not restrict ourselves to the lowest landau level
and if we expand the order parameter over all eigenfunctions with energy $%
\epsilon _{i}<\epsilon _{\ast }$, where the cutting parameter $\epsilon
_{\ast }$ is chosen such that increasing it does not influence the results.
The solid and the dashed curves indicate respectively the stable $L$-states
and the saddle point states. The transition between the $L=1$ state and the $%
L=2$ state is enlarged in the inset. Notice that the stability region of the 
$L$-states is further reduced. Allowing more basis functions in Eq.~(\ref
{vgl4}) does not have a strong influence on the energy of the $L$-states,
e.g. compare the dotted and the solid curves in Fig.~\ref{enR2}(a), but it
considerably decreases the energy of the saddle point between the $L$%
-states. In doing so, it reduces strongly the stability range of the
metastable states, and consequently it reduces the size, i.e. the width in
the magnetic field range, of the hysteresis effect \cite{PRB59}. For
example, we found $(H_{tr}/H_{c2},H_{e}/H_{c2},H_{p}/H_{c2})\approx \left(
1.0,0.52,1.25\right) ,$ $\left( 1.0,0.715,1.245\right) $ and $\left(
1.0,0.73,1.24\right) $ for the $L=0\leftrightarrow L=1$ transition when we
include two, three and an arbitrary number of basis function in Eq.~(\ref
{vgl4}), respectively. Similarly, we found for the $L=1\leftrightarrow L=2$
transition $(H_{tr}/H_{c2},H_{e}/H_{c2},H_{p}/H_{c2})\approx \left(
1.715,1.52,1.81\right) ,$ $\left( 1.715,1.535,1.80\right) ,$ and $\left(
1.715,1.555,1.795\right) $ including two, three and an arbitrary number of
basis functions, respectively. These results clearly show that one has to
exert some caution to cutoff the expansion in Eq.~(\ref{vgl4}) when
calculating the saddle point and thus the energy barriers. Notice that the
expulsion field $H_{e}/H_{c2}$ is most strongly influenced by the number of
terms in Eq.~(\ref{vgl4}).

In Fig.~\ref{barrR2} the transition barriers $U$, i.e. the energy difference
between the saddle point state and the nearby metastable states are plotted.
We show the `exact' numerical results (solid curves) and the results when
including only two values of $l$ with $n=1$, i.e. $\Psi =C_{l_{1}}\varphi
_{l_{1}}+C_{l_{2}}\varphi _{l_{2}}$ (dashed curves). Notice that by
approximating the order parameter better, it substantially lowers the energy
barriers, it increases the expulsion fields $H_{e}$ and lowers the
penetration fields $H_{p}$ slightly. The energy barrier is smaller for
higher $L\rightarrow L+1$ transitions which occur at larger magnetic fields.
The inset shows the barrier between the $L=1$ and the $L=2$ state in more
detail.

The spatial distribution of the superconducting electron density $\left|
\Psi \right| ^{2}$ in the saddle point state corresponding to the transition
from the $L=1$ state to the $L=2$ states is depicted in Figs.~\ref{z12R2}%
(a-d) for the magnetic fields indicated by the open circles in the inset of
Fig.~\ref{barrR2}, i.e. $H_{0}/H_{c2}=1.615$, $1.665$, $1.715$ (i.e. the
maximum of the barrier) and $1.765$, respectively. In Ref.~\cite{PRL83}
similar results were shown for the $L=0\leftrightarrow L=1$ saddle point.
High (low) Cooper-pair density is given by dark (light)\ regions. With
increasing field, one vortex moves from the center to the outer region of
the disk, and the state changes from $L=2$ to $L=1$. This is better
illustrated by the contour plots of the phase of the order parameter which
is shown in Figs.~\ref{z12R2}(e-h) for the same configurations. Along a
closed path, which lies near the edge of the superconductor, the phase
difference $\Delta \varphi $ is always given by $L$ times $2\pi $, with $L$
the vorticity or winding number. Light regions indicate phases $\varphi
\gtrsim 0$ and dark regions $\varphi \lesssim 2\pi $. When encircling the
superconductor near the boundary, we find that the phase difference $\Delta
\varphi $ is equal to $2\times 2\pi $ in Figs.~\ref{z12R2}(e-g) and $\Delta
\varphi =1\times 2\pi $ in Fig.~\ref{z12R2}(h), which means vorticy $L=2$
and $1,$ respectively. At the maximum of the barrier, i.e. when the energy
of state $L=1$ and $L=2$ are identical, the saddle point transits from
vorticity $L=2$ to $L=1$. At this point the Cooper-pair density is zero at
the boundary of the disk which acts as a nucleation center for flux
penetration and expulsion \cite{PRL83}.

\subsection{Large disks: multivortex states}

We consider now a larger superconducting disk with radius $R=4.0\xi $ in
which multivortex states can nucleate in certain magnetic field ranges \cite
{PRL81,Zalalutdinov}. Fig.~\ref{enR4} shows the free energy as a function of
the applied magnetic field $H_{0}$. The energy of the different $L$-states
is given by solid curves when they are in the giant vortex state and by
dashed curves when they are in the multivortex state and the saddle point
states are given by the dotted curves. The open circles give the transition
points between the multivortex state and the giant vortex state for fixed $L$%
. The inset shows the transition barrier $U$ as a function of the applied
field for the different $L\leftrightarrow L+1$ transitions. To distinguish
qualitatively the giant vortex state from the multivortex state for fixed $L$
we considered the value of the Cooper-pair density $\left| \Psi \right| ^{2}$
in the center of \ the disk. Fig.~\ref{denscentR4} shows $\left| \Psi
\right| _{center}^{2}$ which is zero for a giant vortex state and non zero
in the multivortex states when there is no vortex in the center of the disk.
For $R=4.0\xi $ we find only multivortex states for $L=2$, $3$, $4$ and $5$
and the transition from the multivortex state to the giant vortex state
occurs at $H_{MG}/H_{c2}=0.52$, $0.77$, $0.875$ and $0.935$, respectively.
Of course, for $L=1$ there is no distinction between the giant and the
multivortex state. For $L=5$ the spatial distribution of the superconducting
electron density is given in the insets (a-d) of Fig.~\ref{denscentR4} at
the magnetic fields corresponding to the open circles in Fig.~\ref
{denscentR4}, i.e. $H_{0}/H_{c2}=0.8$, $0.85$, $0.9$ and $0.95$,
repectively. In the multivortex state the vortices move towards the center
with increasing magnetic field and at the same time the vortices become
wider, and therefore, the Cooper-pair density in the center decreases until
the axial symmetry is recovered at the transition field $H_{MG}=0.935H_{c2}$.

Next, we will study the energy barriers $U$ in more detail. For a
superconducting disk with radius $R=4.0\xi $ the energy barriers for the
different $L\leftrightarrow L+1$ transitions are shown in the inset of Fig.~%
\ref{enR4}. The height of the energy barrier for the $L\leftrightarrow L+1$
transition decreases with increasing $L$. The difference between penetration
and expulsion field decreases also with increasing $L$. Fig.~\ref{barr45R4}%
(a) shows the free energy of the $L=4$ and the $L=5$ states in more detail
(solid curves for the giant vortex state and dashed curves for the
multivortex state) together with the energy of the saddle point state
between these states (dash-dotted curve). Fig.~\ref{barr45R4}(b) gives the
corresponding energy barrier. The open circles correspond to the transition
from multivortex to giant vortex state. The barrier height is clearly not
influenced by the transition from multivortex to giant vortex state, i.e.
there are no jumps or discontinuities at the transition. The spatial
distribution of the superconducting electron density $\left| \Psi \right|
^{2}$ for this saddle point state is depicted in the insets of Figs.~\ref
{barr45R4}(b) for the configurations indicated by the triangles, i.e. $%
H_{0}/H_{c2}=0.81$, $0.885$, $0.96$ (the barrier maximum) and $1.035$,
respectively. Notice that also in the saddle point the transition between a
multivortex state with $L=5$ and a giant vortex state with $L=4$ is clearly
visible.

Near the maximum of the barrier, the barrier height changes linearly with
magnetic field. Therefore, we can approximate the energy barrier $U$ near
its maximum $U_{max}$ by 
\[
\frac{U}{F_{0}}=\frac{U_{max}}{F_{0}}+\alpha \frac{H-H_{max}}{H_{c2}}\text{ ,%
} 
\]
where the slope $\alpha $ is positive for $H\lesssim H_{max}$ and negative
for $H\gtrsim H_{max}$. In Fig.~\ref{alfaR4} the absolute value of the slope 
$\left| \alpha \right| $ is given as a function of $L$ for $H\lesssim
H_{max} $ by the closed circles and for $H\gtrsim H_{max}$ by the open
circles. The\ absolute value of the slope is different for the left and the
right side of the maximum of the barrier. Notice that for $L=0$ and $L=1$, $%
\left| \alpha \right| $ is larger for $H\lesssim H_{max}$ as compared to $%
H\gtrsim H_{max}$, while for $L>1$ the reverse is true. For increasing $L$
the slope decreases and the behavior could be fitted to 
\[
\left| \alpha _{L\leftrightarrow L+1}(L)\right| =\frac{a+bL}{1+cL}\text{ ,} 
\]
with $a=0.02586,$ $b=-0.00300$ and $c=1.40357$ for $H\lesssim H_{max}$, and $%
a=0.02322,$ $b=-0.00217$ and $c=1.26502$ for $H\gtrsim H_{max}$. These
fitting curves are shown in Fig.~\ref{alfaR4} by the solid line for $%
H\lesssim H_{max}$ and by the dashed line for $H\gtrsim H_{max}$. In the
inset of Fig.~\ref{alfaR4} the maximum of the barrier height $U_{max}$ is
given by the symbols as a function of the vorticity $L$. The barrier height
decreases for increasing vorticity and the behavior could be fitted to 
\[
\frac{U_{max}}{F_{0}}(L)=\frac{a+bL}{1+c\sqrt{L}}\text{ ,} 
\]
with $a=0.07229,$ $b=-0.00791$ and $c=0.48657$, which is shown by the solid
curve.

For larger superconducting disks and higher values of $L$, different
configurations of multivortices can occur with the same vorticity \cite
{Akkermans2,PC332,Geim4}. Fig.~\ref{enR6} shows the free energy as a
function of the applied field for the superconducting states with vorticity $%
L=6$ and $L=7$. For both vorticities two configurations are possible; (i) $L$
vortices on a ring and no vortex in the center (solid curve) and (ii) $L-1$
vortices on a ring and $1$ in the center (dashed curve). The vortex state is
completely determined by the number of vortices in the center $L_{center}$
and the total number of vortices $L$. For this reason we characterized the
states by the indices $\left( L_{center};L\right) $ in Fig.~\ref{enR6}. The
insets (a-d) show the Cooper-pair density at $H_{0}/H_{c2}=0.6$ for the $%
\left( 0;6\right) $ state, the $\left( 1;6\right) $ state, the $\left(
0;7\right) $ state, and the $\left( 1;7\right) $ state, respectively. Notice
further that for such large radius, there is no transition from a
multivortex to a giant vortex state for these values of $L$. In Fig.~\ref
{enR6} the free energy of the saddle point state between the $\left(
1;7\right) $ state and the $\left( 0;6\right) $ state is given by the
dash-dotted curve. This state describes the expulsion of one vortex when the
system transits from the $L=7$ to the $L=6$ configuration and is illustrated
in Figs.~\ref{z67R6} (a-c) where we show the spatial distribution of the
superconducting electron density $\left| \Psi \right| ^{2}$ in the saddle
point state at $H_{0}/H_{c2}=0.5$, $0.6$ and $0.7$, respectively. To transit
from $L=7$ to $L=6$, one vortex on the ring moves towards the outside of the
disk and the vortex in the center takes the free place on the ring. High
(low) Cooper-pair density is given by dark (light) regions.

\section{Superconducting rings}

Now, we will consider superconducting disks with radius $R_{o}$ with a hole
in the center with radius $R_{i}$. For the same reason as in Sec.~III we
make a distinction between small and large systems.

\subsection{Small rings: giant vortex state}

As an example, we consider a superconducting ring with radius $R_{o}=2.0\xi $
and hole radius $R_{i}=1.0\xi $. In Fig.~\ref{enR2r1} the free energy is
shown as a function of the applied magnetic field for the different $L$%
-states (solid curves) together with the saddle point states (dashed
curves). We find giant vortex states with $L=0,1,2,3,4$. Comparing this
result with the result for a disk with radius $R=2.0\xi $, more $L$-states
are possible and the superconducting/normal-transition moves to larger
magnetic fields \cite{PRB61}. The inset shows the energy barrier $U$ for the
transitions between the different $L$-states as a function of the difference
between the applied magnetic field $H_{0}$ and the $L\rightarrow L+1$
transition field $H_{L\rightarrow L+1}$. For increasing $L$, the height of
the energy barrier and the difference between the penetration and the
expulsion field decreases. The energy barrier near its maximum can be
approximated by $U/F_{0}=U_{max}/F_{0}+\alpha (H-H_{max})/H_{c2},$ and we
determined the slope $\alpha _{L\rightarrow L+1}$; $\alpha _{0\rightarrow
1}=-0.8$ for $H\lesssim H_{max}$ and $0.9$ for $H\gtrsim H_{max}$, $\alpha
_{1\rightarrow 2}=-0.6$ for $H\lesssim H_{max}$ and $0.75$ for $H\gtrsim
H_{max}$, $\alpha _{2\rightarrow 3}=-0.3$ for $H\lesssim H_{max}$ and $0.42$
for $H\gtrsim H_{max}$, and $\alpha _{3\rightarrow 4}=-0.013$ for $H\lesssim
H_{max}$ and $0.038$ for $H\gtrsim H_{max}$. The slope decreases again for
increasing $L$ and the absolute value of the slope for $H\lesssim H_{max}$
is smaller than for $H\gtrsim H_{max}$ for every $L$, although the
difference is relatively smaller than in the previous disk case where we
found $\alpha _{0\rightarrow 1}=-0.31$ for $H\lesssim H_{max}$ and $0.44$
for $H\gtrsim H_{max}$, and $\alpha _{1\rightarrow 2}=-0.06$ for $H\lesssim
H_{max}$ and $0.1$ for $H\gtrsim H_{max}$.

Next, we investigate the $2\leftrightarrow 3$ saddle point. At $%
H_{0}/H_{c2}=2.01$ (expulsion field) and $2.535$ (penetration field) the
saddle point state equals the giant vortex states with $L=3$ and $L=2$,
respectively. The transition between these two giant vortex states is
illustrated in Figs.~\ref{z23R2r1}(a-d) which show the spatial distribution
of the superconducting electron density $\left| \Psi \right| ^{2}$
corresponding with the open circles in the inset of Fig.~\ref{enR2r1} at $%
H_{0}/H_{c2}=2.1$, $2.2$, $2.315$ (i.e. the barrier maximum) and $2.4$,
respectively. High (low) density is given by dark (light) regions. With
increasing field one vortex moves from inside the ring, through the
superconducting material, outside the ring. From Fig.~\ref{z23R2r1}(c) one
may infer that the Cooper-pair density is zero along a radial line and that
the vortex is, in fact, a sort of line. That this is not the case can be
seen from the left inset of Fig.~\ref{xminR2r1} which shows the Cooper-pair
density $\left| \psi \right| ^{2}$ along this radial line for $%
H_{0}/H_{c2}=2.315.$ The Cooper-pair density in the superconducting material
is zero only at the center of the vortex which is situated at $x_{min}/\xi
\approx 1.5$ and $\left| \Psi \right| ^{2}$ is very small otherwise, i.e. $%
\left| \Psi \right| ^{2}<0.01$. In Fig.~\ref{xminR2r1} the position of the
vortex, i.e. of $x_{min}$, is shown as a function of the applied field. Over
a narrow field region the vortex moves from the inner boundary towards the
outer boundary. From $H_{0}/H_{c2}=2.01$ to $2.25$ the center of the vortex
is still situated in the hole but the vortex already influences the
superconducting state (see for example Figs.~\ref{z23R2r1}(a,b)). From $%
H_{0}/H_{c2}=2.36$ to $2.535$ the center of the vortex lies outside the
ring, but it has still an influence on the saddle point (see for example
Fig.~\ref{z23R2r1}(d)). In the region $H_{0}/H_{c2}=2.25-2.36$ the center of
the vortex is situated inside the superconductor. This is also illustrated
by the contourplot (right inset of Fig.~\ref{xminR2r1}) for the phase of the
order parameter at $H_{0}/H_{c2}=2.315$, corresponding with the open circle
in Fig.~\ref{xminR2r1}. When encircling the superconductor near the inner
boundary of the ring, we find that the phase difference $\Delta \varphi $ is
equal to $2\times 2\pi $ which implies vorticity $L=2$. When encircling the
superconductor near the outer boundary, we find vorticity $L=3$. If we
choose a path around the vortex (located at $x_{min}$), the phase changes
with $2\pi $ and thus $L=1$. At the transition field $(H_{0}/H_{c2}=2.315)$
the center of the vortex of the saddle point is clearly not situated at the
outer boundary as was the case for superconducting disks (see for example
Figs.~\ref{z12R2}(c,g), \ref{barr45R4}(b), \ref{z67R6}(b)\ and Ref.~\cite
{PRL83}).

To illustrate this more clearly, Figs.~\ref{xminR2}(a,b) show the radial
position of the vortex during the transition between the Meissner state and
the $L=1$ state, and between the $L=1$ state and the $L=2$ state for a
superconducting ring with radius $R_{o}=2.0\xi $ and for several values of
the hole radius, i.e. $R_{i}/\xi =0.0$, $0.5$, $1.0$ and $1.5$. The open
circles indicate the ground state transition fields. Only for the case of
the disk without a hole the center of the vortex at the saddle point occurs
at the outer boundary of the disk for the magnetic field at which the ground
state changes from $L$ to $L+1$. When the disk contains a hole in the center
there are two boundaries and the center of the above vortex is now located
between those two boundaries. For a small hole with radius $R_{i}=0.5\xi $
the position of the vortex can be approached by the arithmetic mean of the
inner and the outer radius, i.e. $x_{min}/\xi \approx (R_{o}+R_{i})/2$, and
for a larger hole with radius $R_{i}=1.5\xi $ by the geometric mean $\sqrt{%
R_{o}R_{i}}$. The transition field increases and the magnetic field range,
over which the transition occurs, decreases with increasing $L$. Notice that
the transition field for the $L=1\leftrightarrow 2$ transition for $%
R_{i}/\xi =0.5$ is larger than the one for $R_{i}/\xi =0.0$ (see Fig.~\ref
{xminR2}(b)), which agrees with Fig.~4 of Ref.~\cite{PRB61}.

\subsection{Large rings: multivortex states}

First, we consider superconducting rings with radius $R_{o}=4.0\xi $ and
hole radius $R_{i}=1.0\xi $. In Fig.~\ref{enR4r1} the free energy is shown
as a function of the applied magnetic field. The different $L$-states are
given by solid curves for giant vortex states and dashed curves for
multivortex states, while the saddle point states are given by the dotted
curves. The open circles correspond to the transition between the
multivortex state and the giant vortex state for fixed $L$. These
transitions occur at $H_{MG}/H_{c2}=0.93$, $1.035$ and $1.14$ for $L=4$, $5$
and $6$, respectively. Notice that for such a small hole in the disk the
maximum number of $L$, i.e. $L=10$, is the same as for the disk case without
a hole (see Fig.~\ref{enR4}). The spatial distribution of the
superconducting electron density $\left| \Psi \right| ^{2}$ is depicted in
the insets (a-c) of Fig.~\ref{enR4r1} for the multivortex state with $L=4$
at $H_{0}/H_{c2}=0.8$, $L=5$ at $H_{0}/H_{c2}=0.9$ and $L=6$ at $%
H_{0}/H_{c2}=1.0$, respectively. High (low) Cooper-pair density is given by
dark (light) regions. Notice that there are always $L-1$ vortices in the
superconducting material and one vortex appears in the hole, i.e. in the
center of the ring.

The energy barriers for the transitions between the different $L$-states are
shown in Fig.~\ref{barrR4r1} as a function of the applied magnetic field. By
comparing this with the energy barriers for a disk with no hole, we see that
the barrier heights and the transition fields are strongly different (see
the inset of Fig.~\ref{enR4}). Therefore we show in the insets of Fig.~\ref
{barrR4r1} the maximum height of the energy barrier $U_{max}$ and the $%
L\leftrightarrow L+1$ transition field $H_{tr}$ as a function of $L$ for
superconducting disks with no hole (squares) and with a hole of radius $%
R_{i}=1.0\xi $ (circles), $2.0\xi $ (triangles) and $3.0\xi $ (stars). In
all cases the height of the energy barrier decreases and the transition
fields increases with increasing $L$. By comparing the situation with no
hole and with a small hole with $R_{i}=1.0\xi $, we see that the barriers
for $L\leq 1$ are higher for the disk with a hole with $R_{i}=1.0\xi $ than
for $R_{i}=0.0\xi $, while they are smaller when $L>1$. Notice also that the
value of the $L\rightarrow L+1$ transition field is sensitive to the
presence of the hole with radius $R_{i}=1.0\xi $ for small $L$ and
insensitive for larger $L$. The reason is that for small $L>0$ such a
central hole has always one vortex localised inside which favors certain
vortex configurations above others, while for larger $L$ in both cases only
giant vortices appear with sizes larger than the hole size and the presence
of the hole no longer matters. For larger holes the energy barrier decreases
more slowly, because the free energy of the different $L$ states shows a
more parabolical type of behavior as a function of the magnetic field. The
transition field has a much smaller dependence on the radius of the hole.
Notice that the transition field for $R_{i}=3.0\xi $ is linear as a function
of $L$ for small holes and $L\leq 9$. This is in good agreement with the
results in the narrow ring limit, where the transition between states with
different vorticity $L$ occurs when the enclosed flux $\phi $ equals $\left(
L+1/2\right) \phi _{o}$ \cite{Thinkham}.

Next, we investigate the saddle point states in these superconducting rings.
We make a distinction between different kinds of saddle point states; i)
between two giant vortex states, ii) between a multivortex and a giant
vortex state, iii) between two multivortex states with the same vorticity in
the hole and different vorticity in the superconducting material, and iv)
between two multivortex states with the same vorticity in the
superconducting material but different vorticity in the hole. The first
saddle point transition was already described for the case of small
superconducting rings (see Figs.~\ref{z23R2r1} and \ref{xminR2r1}). Next, we
study the saddle point state between a multivortex state with $L=5$ and a
giant vortex state with $L=4$ for the previous considered ring with radius $%
R_{o}=4.0\xi $ and hole radius $R_{i}=1.0\xi $. Figs.~\ref{z45R4r1}(a-f)
show the Cooper-pair density for these saddle point states at $%
H_{0}/H_{c2}=0.83$, $0.88$, $0.93$, $0.965$ (i.e. the barrier maximum), $%
1.03 $ and $1.06$, respectively. High (low) Cooper-pair density is given by
dark (light) regions. For increasing field one vortex moves to the outer
boundary, while the others move to the center of the ring where they create
a giant vortex state. Remark that the giant vortex state is larger than the
hole and therefore it is partially situated in the superconductor itself.

To study saddle point transitions between different multivortex states we
have to increase the radius of the ring to favour the multivortex states.
Therefore, we consider a ring with radius $R_{o}=6.0\xi $ and hole radius $%
R_{i}=2.0\xi $. Fig.~\ref{enR6r2} shows the free energy of multivortex
states with $L=8$ and $L=9$. In both cases 3 vortices are trapped in the
hole. The lower insets show the spatial distribution of the superconducting
electron density $\left| \Psi \right| ^{2}$ at the transition field $%
H_{0}/H_{c2}=0.695$ for $L=8$ and $L=9$. It is clear that there are only 5
and 6 vortices in the superconducting material, respectively. The free
energy of these multivortex states is shown by solid curves, while the
saddle point energy between these states is given by the dashed curve.
Notice further, that there is no transition from the multivortex states to
the giant vortex states with $L=8$ and $9$ as long as these states are
stable. The spatial distribution of the superconducting electron density $%
\left| \Psi \right| ^{2}$ for this saddle point state is depicted in the
upper insets at the magnetic fields $H_{0}/H_{c2}=0.63$, $0.695$ (the
barrier maximum) and $0.76$, respectively. For increasing field one vortex
moves from the superconducting material to the outer boundary and hence the
vorticity changes from $L=9$ to $L=8.$ Notice that the vorticity of the
interior boundary of the ring does not change.

The fourth type of saddle point state to discuss is the $L\rightarrow L+1$
transition between two multivortex states with the same vorticity in the
superconducting material but with a different vorticity in the hole. For $%
R_{o}/\xi =4$ and $R_{o}/\xi =6$ we did not find such transitions regardless
of the hole radius. This means that at least for these radii there is no
transition between such states which describes the motion of one vortex from
the hole through the superconducting material towards the outer insulator.

Finally, we investigated the influence of the hole radius on the barrier for
a fixed outer ring radius. Fig.~\ref{barrhole}(a) shows the maximum barrier
height, i.e. the barrier height at the thermodynamic equilibrium $%
L\rightarrow L+1$ transition, as a function of the hole radius $R_{i}$ for a
ring with radius $R_{o}=4.0\xi $ for the transition between the Meissner
state and the $L=1$ state (solid curve) and for the transition between the $%
L=1$ and the $L=2$ state (dashed curve). For increasing hole radius, the
barrier height of the first transition rapidly increases in the range $%
R_{i}=0.1\xi $ to $R_{i}=1.5\xi $ and decreases slowly afterwards. For a
superconducting disk with radius $R_{o}=4.0\xi $ with a hole in the center
with radius $R_{i}=1.5\xi $ the maximum barrier height for the $0\rightarrow
1$ transition is twice as large as for a superconducting disk without a
hole. The barrier height of the second transition first decreases, then
rapidly increases in the range $R_{i}=0.6\xi $ to $R_{i}=2.5\xi $ and then
slowly decreases again. In this case the maximum barrier height for a
superconducting disk with a hole with radius $R_{i}=2.5\xi $ is three times
as large as for a superconducting disk without a hole. Hence, changing the
hole radius strongly influences the maximum height of the barrier. In Fig.~%
\ref{barrhole}(b) we plot the characteristic magnetic fields of the barrier
as a function of the hole radius, i.e. the transition magnetic field $H_{tr}$%
, the expulsion magnetic field $H_{e}$\ and the penetration magnetic field $%
H_{p}$, for the $0\rightarrow 1$ transition by solid curves and for the $%
1\rightarrow 2$ transition by the dashed curve. For the $0\rightarrow 1$
transition the characteristic magnetic fields decrease with increasing hole
radius. For the $1\rightarrow 2$ transition the characteristic magnetic
fields first increase to a maximum and then decrease. This behaviour was
described and explained in our previous paper (see e.g. Fig.~17 of Ref.~\cite
{PRB61}). Notice that the position of the minimum in $U_{max}$ coincides
with the position of the maximum in $H_{tr}$.

\section{Conclusions}

We studied the saddle points for transitions between different vortex states
for thin superconducting disks and rings. A distinction was made between
small systems where the confinement effects dominate and larger systems
where multivortices can nucleate for certain magnetic fields. At the
entrance of the vortex into the superconducting material the superconducting
density becomes zero at a certain point at the edge of the disk or ring.
Such a zero in the order parameter acts as a phase slip center which allows
the vorticity to increase with one unit. For the case of the ring the vortex
may enter (or exit) the superconducting material from the inner boundary or
from the outer boundary of the ring.

We studied the transition between two giant vortex states with different
vorticity $L$. One vortex moves through the superconducting material to the
center of the disk or to the hole. During the transition the position of
this vortex in the superconductor can be determined very precisely, because
the Cooper-pair density is exactly zero in the center of this vortex. The
transition between a multivortex state and a giant vortex state with
different vorticity $L$ is also described. One vortex leaves (enters) the
superconductor while the other vortices move towards (away from) the center
of the disk. For large enough disk/ring radii, we calculated the transition
between two multivortex states. We found such transitions between two
multivortex states with different vorticity $L$ in the superconductor but
with the same vorticity in the center/hole. One vortex enters/leaves the
superconductor while the other vortices rearrange themselves. Transitions
between different multivortex states with the same vorticity in the
superconducting material but different vorticity in the hole were not found
for the considered ring configurations, which means that transitions between
such states do not occur in these particular cases.

The maximum height of the energy barrier always decreases for increasing $L$%
. Near the maximum, the barrier height decreases linearly and its slope at
the left side $\left( H\lesssim H_{max}\right) $ of the maximum is not equal
to the slope at the right side $\left( H\gtrsim H_{max}\right) $. The
barrier shape and height strongly depend on the radius of the hole in the
center of the disk.

\section{Acknowledgments}

This work was supported by the Flemish Science Foundation (FWO-Vl), the
''Onderzoeksraad van de Universiteit Antwerpen'', the ''Interuniversity
Poles of Attraction Program - Belgian State, Prime Minister's Office -
Federal Office for Scientific, Technical and Cultural Affairs'', and the
European ESF-Vortex Matter. Discussions with S. Yampolskii are gratefully
acknowledged.

\bigskip

\begin{figure}[tbp]
\caption{The energy of the minima in the free energy $F$ and the energy of
the saddle points as a function of the applied magnetic field $H_{0}$ for a
superconducting disk with radius $R=2.0\protect\xi .$ (a) When only one term
is included, i.e. $(n,l)=(1,l)$ (dotted curve), when two $l$-values are
included, i.e. $(n,l_{1})$ and $(n,l_{2})$ (solid curves), with the
corresponding energy of the saddle point (dashed curves); (b) The giant
vortex energy (solid curves) and the saddle point energy (dashed curves)
when an arbitrary large number of terms are included. The free energy is
scaled with the condensation energy $F_{0}=\protect\alpha ^{2}\protect\pi
R^{2}d/2\protect\beta $.}
\label{enR2}
\end{figure}

\begin{figure}[tbp]
\caption{The transition barrier $U$ for transitions between different $L$%
-states for a superconducting disk with radius $R=2.0\protect\xi $ when
taking into account only two values of $l$ (dashed curves) and for the
numerical `exact' result (solid curves). The inset shows the second barrier
in more detail.}
\label{barrR2}
\end{figure}

\begin{figure}[tbp]
\caption{The spatial distribution of the superconducting electron density $%
\left| \Psi \right| ^{2}$ (a-d) and the phase of the order parameter (e-h)
in the saddle point state corresponding to the transition from the $L=1$
state to the $L=2$ states in a superconducting disk with radius $R=2.0%
\protect\xi $ for the magnetic fields indicated by the open circles in the
inset of Fig.~\ref{barrR2}; $H_{0}/H_{c2}=1.615$ (a,e), $1.665$ (b,f), $%
1.715 $ (i.e. the maximum of the barrier) (c,g) and $1.765$ (d,h). High
Cooper-pair density is given by dark regions, low Cooper-pair density by
light regions. Phases $\protect\varphi \gtrsim 0$ are given by light regions
and $\protect\varphi \lesssim 2\protect\pi $ by dark regions. }
\label{z12R2}
\end{figure}

\begin{figure}[tbp]
\caption{The free energy as a function of the applied magnetic field $H_{0}$%
. The different $L$-states are given by solid curves when in the giant
vortex states and by dashed curves when in the multivortex states, while the
saddle point states are given by the dotted curves. The open circles
correspond to the transitions between the multivortex state and the giant
vortex state for fixed $L$. The inset shows the transition barrier $U$ as a
function of the applied field for the different $L\leftrightarrow L+1$
transitions.}
\label{enR4}
\end{figure}

\begin{figure}[tbp]
\caption{The Cooper-pair density $\left| \Psi \right| ^{2}$ in the center of
\ the disk with radius $R=4.0\protect\xi $ for $L=2$, $3$, $4$ and $5$. The
insets (a-d) show the spatial distribution of the superconducting electron
density for $L=5$ at the magnetic fields corresponding to the open circles;
i.e. $H_{0}/H_{c2}=0.8$, $0.85$, $0.9$ and $0.95$, repectively. The
transition from multivortex state to giant vortex state occurs at the
transition field $H_{MG}$.}
\label{denscentR4}
\end{figure}

\begin{figure}[tbp]
\caption{(a) The free energy of the $L=4$ and the $L=5$ states (solid curves
for giant vortex states and dashed curves for multivortex states) and the
saddle point states between these states (dash-dotted curve) for a
superconducting disk with radius $R=4.0\protect\xi $; and (b) the energy
barrier corresponding with this transition. The open circles correspond with
the transition from multivortex to giant vortex state for fixed $L$. The
insets show the spatial distribution of the superconducting electron density 
$\left| \Psi \right| ^{2}$ for the saddle point states indicated by
triangles, i.e. at the magnetic fields $H_{0}/H_{c2}=0.81$, $0.885$, $0.96$
(the barrier maximum) and $1.035$. It is the transition between a
multivortex state with $L=5$ and a giant vortex state with $L=4$.}
\label{barr45R4}
\end{figure}

\begin{figure}[tbp]
\caption{The absolute value of the slope $\left| \protect\alpha \right| $ of
the energy barrier for a superconducting disk with $R=4.0\protect\xi $ as a
function of $L$ for $H\lesssim H_{max}$ (closed circles) and for $H\gtrsim
H_{max}$ (open circles). The inset shows the maximum barrier height as a
function of the vorticity $L$. The solid and dashed curves are the results
of a fit. }
\label{alfaR4}
\end{figure}

\begin{figure}[tbp]
\caption{The free energy $F$ as a function of the applied magnetic field $%
H_{0}$ of the $(0;6)$ and the $(0;7)$ state (solid curves), the $(1;6)$ and
the $(1;7)$ state (dashed curves), and the saddle point state (dash-dotted
curve) between the $(1;7)$ and $(0;6)$ state for a superconducting disk with
radius $R=6.0\protect\xi $. The insets show the Cooper-pair density of the $%
(0;6)$ state (a), the $(1;6)$ state (b), the $(0;7)$ state (c), and the $%
(1;7)$ state (d) at the transition field $H_{0}/H_{c2}=0.6$.}
\label{enR6}
\end{figure}

\begin{figure}[tbp]
\caption{The spatial distribution of the superconducting electron density $%
\left| \Psi \right| ^{2}$ for the transition between the $L=6$ state and the 
$L=7$ state for a superconducting disk with radius $R/\protect\xi =6.0$ at
the applied magnetic fields $H_{0}/H_{c2}=0.5$ (a), $0.6$ (b) and $0.7$ (c).}
\label{z67R6}
\end{figure}

\begin{figure}[tbp]
\caption{The free energy for a superconducting ring with radius $R_{o}=2.0%
\protect\xi $ and hole radius $R_{i}=1.0\protect\xi $ as a function of the
applied magnetic field for the different giant vortex states (solid curves)
and for the saddle point states (dashed curves). The inset shows the energy
barrier $U$ for the transitions between different $L$-states as a function
of the difference between the applied magnetic field $H_{0}$ and the $%
L\rightarrow L+1$ transition field $H_{L\rightarrow L+1}$.}
\label{enR2r1}
\end{figure}

\begin{figure}[tbp]
\caption{The spatial distribution of the superconducting electron density $%
\left| \Psi \right| ^{2}$ of the transition between the giant vortex states
with $L=2$ and $L=3$ for a superconducting ring with $R_{o}=2.0\protect\xi $
and $R_{i}=1.0\protect\xi $ at $H_{0}/H_{c2}=2.1$ (a), $2.2$ (b), $2.315$
(c) and $2.4$ (d). High density is given by dark regions and low density by
light regions.}
\label{z23R2r1}
\end{figure}

\begin{figure}[tbp]
\caption{The radial position of the vortex in the saddle point for the $%
2\leftrightarrow 3$ transition through the superconductor with radius $%
R_{o}=2.0\protect\xi $ and $R_{i}=1.0\protect\xi $. The left inset shows the
Cooper-pair density along de $x$-direction at $H_{0}/H_{c2}=2.315,$ and the
right inset is a contour plot of the phase of the order parameter at $%
H_{0}/H_{c2}=2.315$.}
\label{xminR2r1}
\end{figure}

\begin{figure}[tbp]
\caption{The radial position of the vortex for (a) the $0\leftrightarrow 1$
and (b) the $1\leftrightarrow 2$ saddle point transition as a function of
the applied magnetic field for a superconducting ring with radius $R_{o}=2.0%
\protect\xi $ and $R_{i}=0.0$, $0.5$, $1.0$ and $1.5\protect\xi $. The open
circles indicate the transition fields.}
\label{xminR2}
\end{figure}

\begin{figure}[tbp]
\caption{The free energy for a superconducting ring with $R_{o}=4.0\protect%
\xi $ and $R_{i}=1.0\protect\xi $\ as a function of the applied magnetic
field for the different $L$-states (solid curves for giant vortex states and
dashed curves for multivortex states), and the saddle point states (dotted
curves). The open circles correspond to the transition between the
multivortex state and the giant vortex state for fixed $L$. The inset shows
the spatial distribution of the superconducting electron density $\left|
\Psi \right| ^{2}$ for the multivortex state with $L=4$ at $H_{0}/H_{c2}=0.8$
(a), $L=5$ at $H_{0}/H_{c2}=0.9$ (b) and $L=6$ at $H_{0}/H_{c2}=1.0$ (c).
High Cooper-pair density is given by dark regions, low Cooper-pair density
by light regions.}
\label{enR4r1}
\end{figure}

\begin{figure}[tbp]
\caption{The energy barrier $U$ for the transitions between the different $L$%
-states in a superconducting ring with $R_{o}=4.0\protect\xi $ and $R_{i}=1.0%
\protect\xi $ as a function of the applied magnetic field. The insets show
the maximum height of the energy barrier $U_{max}$ and the transition field $%
H_{tr}$ as a function of $L$ for rings with $R_{o}=4.0\protect\xi $ and $%
R_{i}=0.0$, $1.0$, $2.0$, and $3.0\protect\xi $.}
\label{barrR4r1}
\end{figure}

\begin{figure}[tbp]
\caption{The Cooper-pair density for the saddle point state transition
between a multivortex state with $L=5$ and a giant vortex state with $L=4$
at $H_{0}/H_{c2}=0.83$ (a), $0.88$ (b), $0.93$ (c), $0.965$ (d), $1.03$ (e)
and $1.06$ (f). High Cooper-pair density is given by dark regions, low
Cooper-pair density by light regions.}
\label{z45R4r1}
\end{figure}

\begin{figure}[tbp]
\caption{The free energy of the multivortex states with $L=8$ and $L=9$
(solid curves) and the saddle point state (dashed curves) between these
multivortex states for a superconducting ring with $R_{o}=6.0\protect\xi $
and $R_{i}=2.0\protect\xi $ as a function of the applied magnetic field. The
lower insets show the spatial distribution of the superconducting electron
density $\left| \Psi \right| ^{2}$ at the transition field $%
H_{0}/H_{c2}=0.695$ for $L=8$ and $L=9$. The upper insets show the spatial
distribution of the superconducting electron density $\left| \Psi \right|
^{2}$ for the saddle point states indicated by the open circles, i.e. at $%
H_{0}/H_{c2}=0.63$ (a), $0.695$ (b) and $0.76$ (c). High Cooper-pair density
is given by dark regions, low Cooper-pair density by light regions.}
\label{enR6r2}
\end{figure}

\begin{figure}[tbp]
\caption{(a) The maximum barrier height as a function of the hole radius $%
R_{i}$ for a ring with radius $R_{o}=4.0\protect\xi $ for the transition
between the Meissner state and the $L=1$ state (solid curve) and the
transition between the $L=1$ state and the $L=2$ state (dashed curve); and
(b) the transition magnetic field $H_{tr}$, the expulsion magnetic field $%
H_{e}$ and the penetration magnetic field $H_{p}$ as a function of the hole
radius $R_{i}$ for the transition between the Meissner state and the $L=1$
state (solid curve) and the transition between the $L=1$ state and the $L=2$
state (dashed curve).}
\label{barrhole}
\end{figure}

\end{document}